\documentstyle[aps,epsf,pre,floats]{revtex}

\begin{document}
\twocolumn[{\hsize\textwidth\columnwidth\hsize\csname
@twocolumnfalse\endcsname

\title{
\draft Scaling Concepts in Periodically
Modulated Noisy Systems
}
\author{
J. M. G. Vilar and J. M. Rub\'{\i}
}
\address{
Departament de F\'{\i}sica Fonamental, Facultat de
F\'{\i}sica, Universitat de Barcelona,
Diagonal 647, E-08028 Barcelona, Spain\\
\date{\today}
} 
\maketitle
\widetext
\begin{abstract}
\leftskip 54.8pt
\rightskip 54.8pt

We show that scaling arguments are very useful to analyze 
the dynamics of periodically modulated noisy systems. Information
about the behavior of the relevant quantities, such as the 
signal-to-noise ratio, upon variations of the noise level,
can be obtained by analyzing the symmetries and invariances
of the system.
In this way, it is possible to predict diverse physical manifestations
of the cooperative behavior between noise and input signal,
as for instance stochastic resonance, spatiotemporal stochastic resonance,
and stochastic multiresonance.

\end{abstract}

}]
\narrowtext

\section{Introduction}

The application of scaling concepts in Statistical Physics has revealed
their great usefulness to obtain relevant information about statical and
dynamical quantities of the systems with little effort, without entering
into complex calculations \cite{kada}.
These techniques, which constitute simple
manipulations allowing us to relate different quantities and exponents,
have been employed in many classical branches of Statistical Physics,
among others: critical phenomena \cite{bin,stan},
hydrodynamics \cite{frish,dau}, polymer physics \cite{dg} and
non-linear physics \cite{nico,feig}. More recently,
scaling concepts have also been used
in fields as, to mention just a few, growth phenomena \cite{bara},
fractures \cite{her}, and economy \cite{bou}.
Frequently, the existence of a scaling regime can be made evident by means of
dimensional analysis or by considering the invariances of the equations
involving the relevant quantities of the system.
In this regard, the symmetries present in the underlying equations
arise naturally in their physical realization.

Our purpose in this paper is to show how scaling arguments can also be
applied to analyze the dynamics of
a wide class of systems exhibiting, as a main characteristic,
a dynamical evolution which  is
periodically modulated  when the system is
affected by  noise.
In particular, we will focus on the possibility that the
response of the system  may be enhanced
by the addition of noise. 
This phenomenon, known as stochastic resonance (SR)
\cite{Benzi,tri,Ma1,gam,Ma2,neu1,JSP,neu2,%
Moss,Wies,Wiese,array,thre1,grifo,phi4,prl3,prl2,prl1,thre2,BV},
shows a constructive role of noise, sometimes considered
counterintuitive, since the addition of noise is able to
decrease the randomness of the system.
The importance and interest of SR has been revealed
by the great number of situations in which it has been predicted
and/or observed; ranging from systems as simple as a single dipole
\cite{Agus}
to systems exhibiting certain degree of complexity as
neural tissues \cite{nem} or pattern-forming systems \cite{prl3}.
\begin{figure}[th]

\hfill{\large \it a\ }
\centerline{
\epsfxsize=8cm 
\epsffile{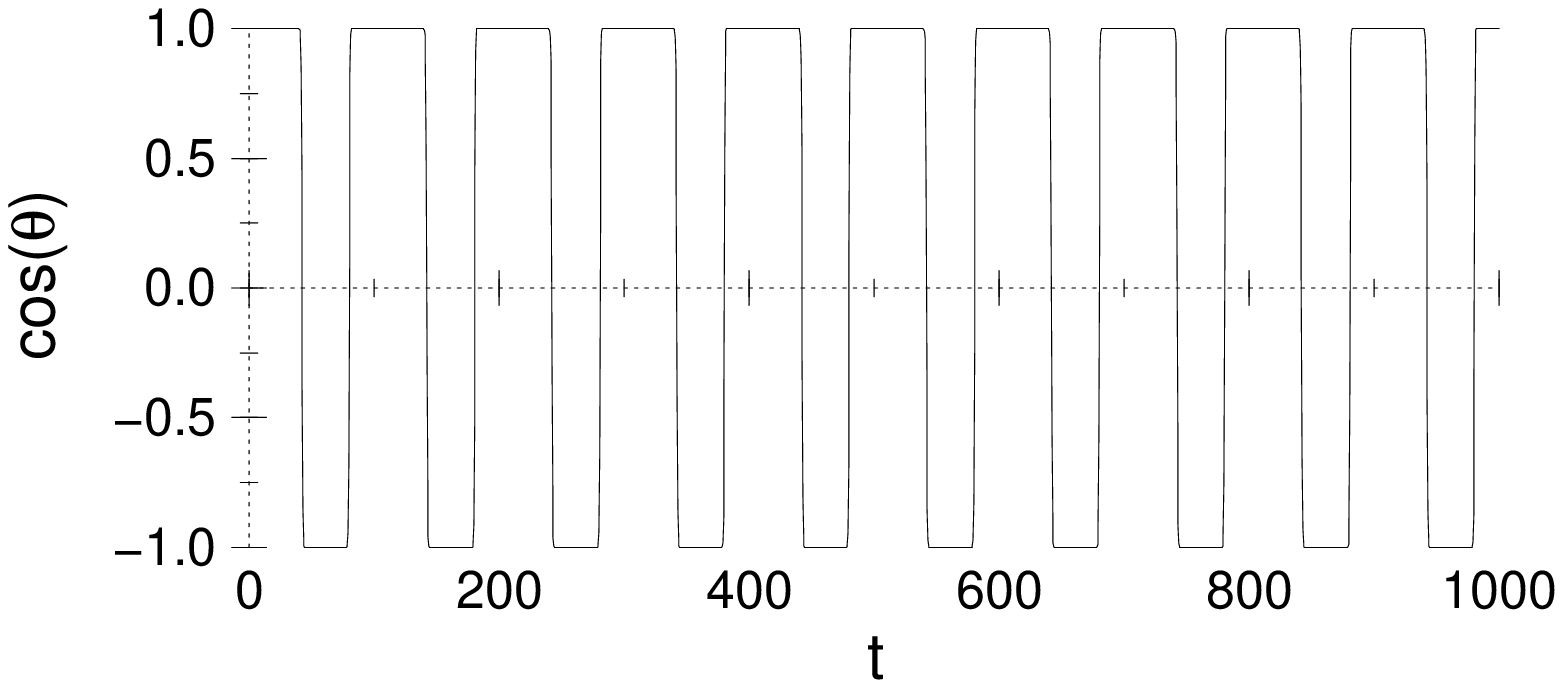}
}

\hfill{\large \it b\ }
\centerline{
\epsfxsize=8cm 
\epsffile{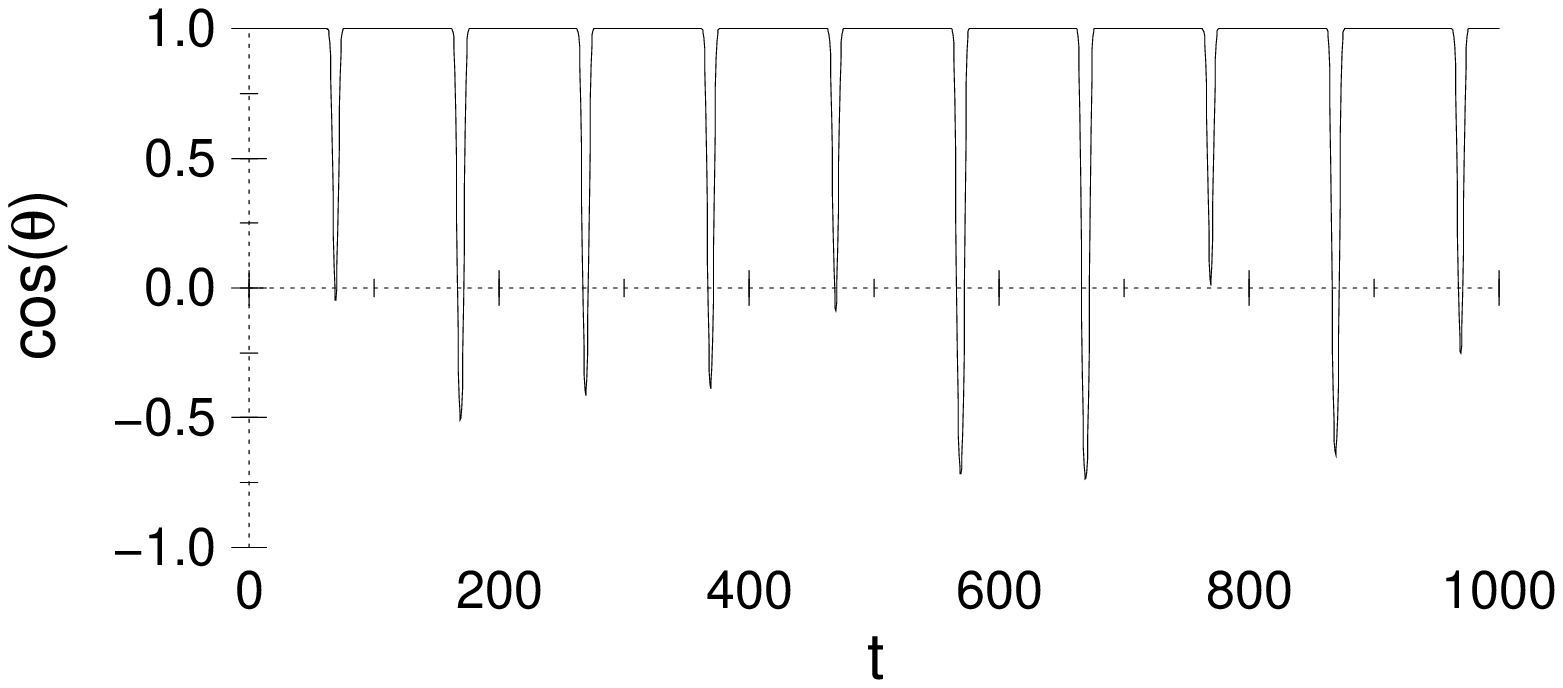}
}

\caption[a]{\label{dipo}
Time evolution of $\cos(\theta)$ (see \cite{dipeq}) for $\alpha=3$,
$k=1$, $\omega_0/2\pi=0.1$ and the noise levels:
(a) $D=10^{-10}$ and (b) $D=10^{-44}$.
}
\end{figure}

To illustrate the essentials of the
phenomenon we will show an instance of how noise,
under some circumstances, is able to increase the  order
of a system.
In Fig. \ref{dipo} we have displayed the 
time evolution corresponding to
the  relaxational  dynamics of a single  dipole \cite{Agus}
characterized by the angle $\theta$
between the direction of the dipole moment  and the direction
of an  external  applied field (see \cite{dipeq}).
Although apparently Fig. \ref{dipo}a looks deterministic, i.e. periodic,
noise is responsible for this order, since  the  behavior becomes
random when noise is sufficiently decreased, as shown in Fig.\ref{dipo}b.
This figure, then, makes the fact that noise is not always
a source of disorder evident.

Along this paper we present a methodology based upon general scaling
arguments which enables one to predict the appearance of such an ordered
behavior due to the presence of noise.
To this end we have organized the paper in the following way.
In Sec. \ref{sec2}
we present results for the simple case of an Ornstein-Uhlenbeck
process. Sec. \ref{sec3} deals with the application of scaling concepts to
systems described by scale-invariant potentials. In particular, we show
the appearance of stochastic resonance for monostable potentials
(Sec. \ref{sec3b}) and analyze how to deal with
spatially extended systems (Sec. \ref{sec3a}). In
Sec. \ref{sec4} we consider potentials which are not scale invariant.  We
discuss how the terms that break the scale invariance 
can be treated perturbatively.
In this context, we also study the phenomenon of spatiotemporal
stochastic resonance,
occurring in Rayleigh-B\'enard convection, and in general in
pattern-forming systems,  through the scaling 
of the Swift-Hohenberg equation.
In Sec. \ref{sec5} we show how the analysis of the discrete symmetries of a
particular class of dynamical systems allows us to predict the appearance
of multiple peaks in the signal-to-noise ratio,
giving rise to stochastic multiresonance.
Finally, in Sec. \ref{sec6}
we summarize our main results  and outline further applications of
scaling concepts to this field.

\section{The periodically modulated Ornstein-Uhlenbeck process}
\label{sec2}

Let us first discuss in detail one of the most
simplest cases which can be treated exactly by only
using dimensional analysis.
Its dynamics
is described by an Ornstein-Uhlenbeck
process, where the input signal modulates the strength of the
force in the following way:
\begin{equation} \label{model1}
{dx \over dt} = -\kappa[1+\alpha\sin(\omega_0 t)]x + \xi(t) \;\; .
\end{equation}
Here $\kappa$, $\alpha$, and $\omega_0$
are constants and  $\xi(t)$ is Gaussian white
noise with zero mean and correlation function
$\left< \xi(t) \xi(t+ \tau) \right> = 2D \delta(\tau)$,
defining the noise level $D$.
In spite of its simplicity,
the previous model encompasses many physical situations of
interest since the motion around a minimum in a force field
whose intensity varies periodically in time can be described
by Eq. (\ref{model1}).

For the sake of generality we will consider that the
system is described by means of a function $v(t)$ of 
the dynamical variable $x$; i. e., $v(t)\equiv v[x(t)]$.
The effect of the force may be
analyzed by the averaged power spectrum 
\begin{equation} \label{ps}
P(\omega)={\omega_0 \over 2\pi}
\int_0^{2\pi/\omega_0}dt \int_{-\infty}^\infty
\left<v(t)v(t+\tau)\right>e^{-i\omega \tau}d\tau \;\; .
\end{equation}
To this end we will assume that it consists of a delta
function centered at the frequency $\omega_0$ plus a function $Q(\omega)$
which is smooth in the neighborhood of $\omega_0$ and is given by
\begin{equation} \label{ps2}
P(\omega)
=Q(\omega)+S(\omega_0)\delta(\omega-\omega_0) \;\;.
\end{equation}
We have expressed the power spectrum in this form since we are interested
in the behavior of the system in the range of frequencies close to the
frequency of the input signal.

The power spectrum, as expressed previously,
explicitly shows the intensity of the deterministic component
of the system or output signal, $S(\omega_0)$,  and the stochastic
component or output noise, $Q(\omega)$. The SNR, defined
as the ratio between the signal and noise,
\begin{equation}
\label{SNRdef}
\mbox{SNR}=S(\omega_0) / Q(\omega_0) \;,
\end{equation}
then indicates the order present in the
system.

Let us now assume the explicit form for the output of the system,
$v(x)=\vert x \vert^\beta$, where $\beta$ is a constant.
Considerations  based upon dimensional analysis
enable us to rewrite the averaged power spectrum as
\begin{eqnarray} \label{mastereq}
P(\omega,D,\kappa,\alpha,\omega_0,\beta)
&=& {1 \over \kappa } \left({D \over \kappa }\right)^\beta
q({\omega / \omega_0},{\kappa  / \omega_0},\alpha,\beta) \nonumber \\
&+& \left({D \over \kappa }\right)^\beta
s({\kappa  / \omega_0},\alpha,\beta)
\delta\left(1-{\omega\over\omega_0}\right) \;\;,
\nonumber \\
& &
\end{eqnarray}
where $q({\omega / \omega_0},{\kappa  / \omega_0},\alpha)$
and $s({\kappa  / \omega_0},\alpha)$ are dimensionless functions.
Note that the previous equation is an exact expression
for the power spectrum since it does not involve any
approximation.

From Eq. (\ref{mastereq}) we can identify the expression
for the output signal
\begin{equation}
\label{psou}
S(\omega_0)=\left({D \over \kappa }\right)^\beta
s({\kappa  / \omega_0},\alpha,\beta) \;\; .
\end{equation}
Thus, we have easily obtained the exact dependence of the output signal
with the noise level. A remarkable aspect that must be emphasized
is the fact that the output signal depends on the quantity
we measure and consequently on the exponent $\beta$ \cite{preR}.
In this respect, inspection of Eq. (\ref{psou}) reveals the
presence of 
three qualitative different situations.
For $\beta>0$ the signal diverges when  the noise level $D$
goes to infinity, whereas for $\beta<0$ the signal
diverges when $D$ goes to zero.
Even more interesting is the limit case $\beta=0$, in which
the signal does not depend on the noise level.

\begin{figure}[th]

\hfill{\large \it a\ }
\centerline{
\epsfxsize=8cm 
\epsffile{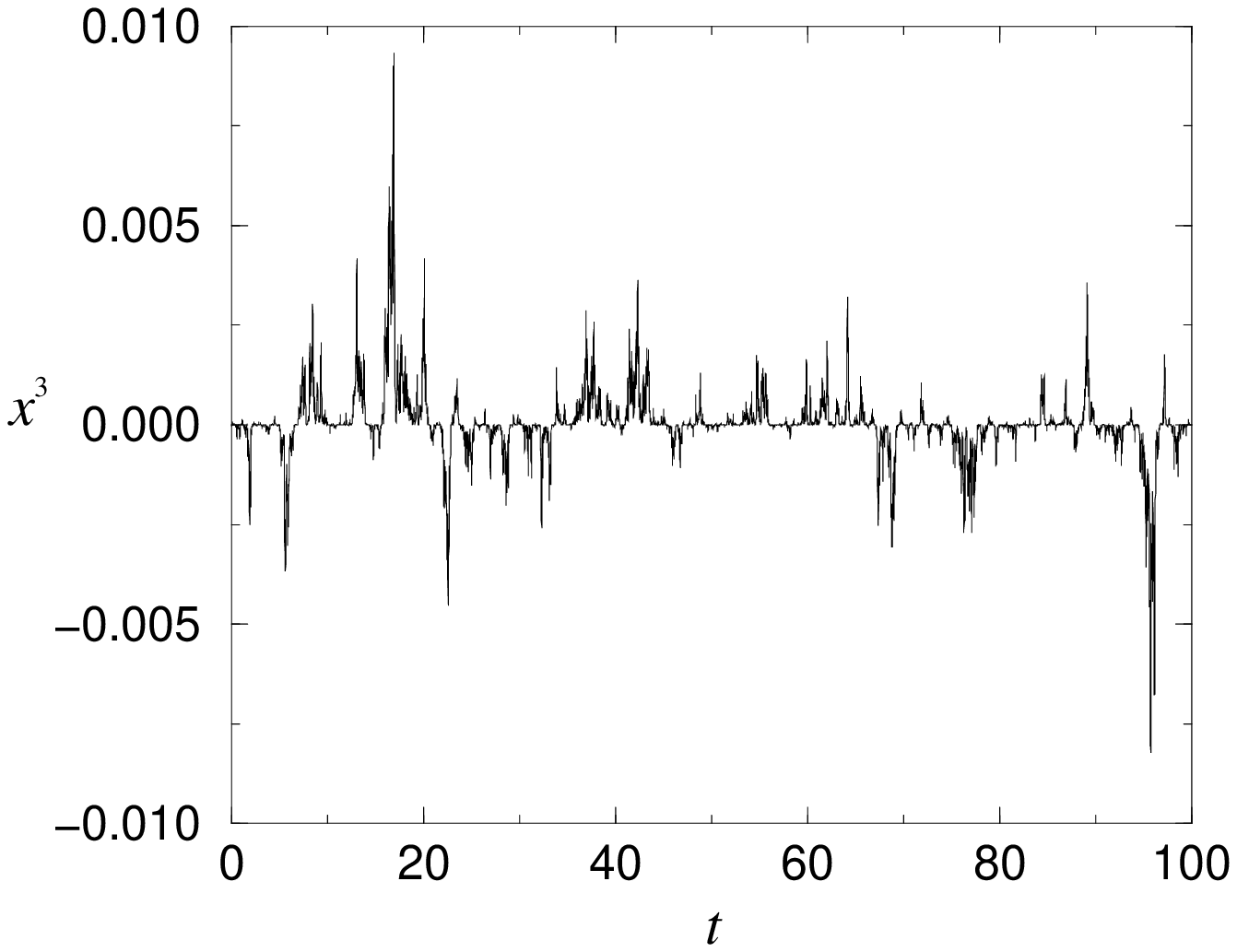}
}

\hfill{\large \it b\ }
\centerline{
\epsfxsize=8cm 
\epsffile{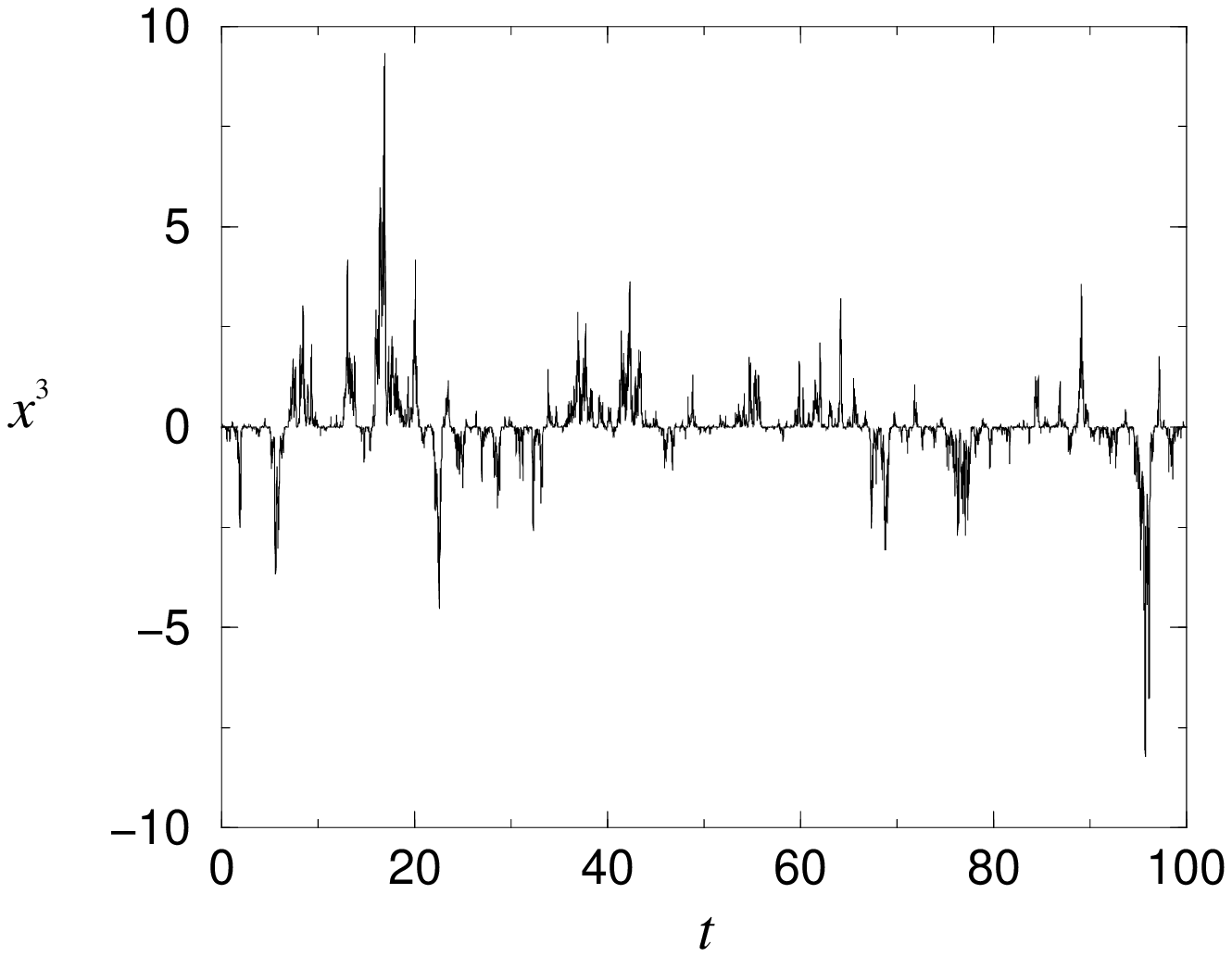}
}

\caption[f]{\label{fig2}
Time evolution of $x^3$ ($\kappa =1$, $\alpha=0.5$, and $\omega_0/2\pi=0.1$)
for the noise levels (a) $D=0.01$ and (b) $D=1$.
}
\end{figure}

The expression for the SNR
straightforwardly follows from Eq. (\ref{mastereq})
\begin{equation}
\mbox{SNR}= \kappa  {s({\kappa  / \omega_0},\alpha,\beta)\over
q({\omega / \omega_0},{\kappa  / \omega_0},\alpha,\beta)} \;\; .
\end{equation}
In contrast to the case for the signal, 
this result does not depend on the noise level thus 
indicating that the system is insensitive to the noise.
No matter the noise intensity, the SNR has always the same value
despite the fact that
signal is a monotonic increasing or decreasing function of
the noise level.

For further illustration of the previous results
we have depicted in Fig. \ref{fig2}.
the time evolution of the output of the system
when $v(x)=x^3$, for two values of
the noise level. In both cases we have used the same
realization of the noise.
In the figure we can see how the noise only affects the system
by changing its characteristic scales.
Therefore, the dependence of the quantities
of interest with the noise level can  also be obtained from
the inspection of 
the invariance properties of the equations under scale transformations.
Thus when rescaling the noise level,  $x$, and $t$ in the
following way:
\begin{eqnarray}
D &\rightarrow D^{\prime} \equiv& bD \;\;, \nonumber\\
x &\rightarrow x^{\prime} \equiv& b^{\gamma_1}x \;\;,\nonumber \\
\label{res3} t &\rightarrow t^{\prime} \equiv& b^{\gamma_2}t \;\;,
\end{eqnarray}
Eq. (\ref{model1}) must be independent of $b$ for the
adequate values of the exponents $\gamma_1$
and $\gamma_2$. Consequently, by substituting
Eqs. (\ref{res3}) into
Eq. (\ref{model1}) we obtain
\begin{equation} \label{model1b}
b^{\gamma_1-\gamma_2}{dx \over dt} =
-\kappa[1+\alpha\sin(\omega_0 b^{\gamma_2}t)]b^{\gamma_1}x
+ b^{1/2-\gamma_2/2}\xi(t) \;\; ,
\end{equation}
which is left unchanged when  $\gamma_1=1/2$ and $\gamma_2=0$.
Since the power spectrum, given in Eq. (\ref{ps}), transforms under the
scaling (\ref{res3}) as
\begin{equation} \label{ps3}
P^{\prime}(\omega^{\prime})
=b^{2\beta\gamma_1+\gamma_2}Q(\omega)
+b^{2\beta\gamma_1}S(\omega_0)
\delta(\omega^{\prime}-\omega_0^{\prime}) \;, 
\end{equation}
then, the output signal and SNR scale with the noise level
as $S(\omega_0) \sim D^{\beta}$
and $\mbox{SNR} \sim D^0$.

\section{Scale invariant potentials}
\label{sec3}

In this section we will discuss two different cases which
share as a common characteristic the scale invariance of the potential.
In Sec. \ref{sec3b} we will
deal with nonlinear systems by using the analysis
introduced in the previous section, 
whereas in Sec. \ref{sec3a} 
we will extend
those ideas to study spatial effects in a linear system.

\subsection{Nonlinear systems}
\label{sec3b}

The previous results we have obtained are exact since no
approximations has been made. In order to 
analyze more interesting situations a similar
scheme can be followed together with some assumptions
about the behavior of the quantities of interest. Below we will
discuss some of such situations.

The class of systems we will now consider are described by only one
relevant degree of freedom whose dynamics is governed by the
following Langevin equation
\begin{equation}
\label{ourmodel}
{dx \over dt} = -\kappa[1+\alpha\sin(\omega_0 t)]x^{1+2n} + \xi(t) \;\; , 
\end{equation}
where  $\kappa$ and $\alpha$
($<1$) are constants, $n$ is an integer number and $\xi(t)$ is Gaussian
white noise with zero mean and correlation function
$\left< \xi(t) \xi(t+\tau) \right> = 2D \delta(\tau)$.

From its definition [Eq. (\ref{SNRdef})]
we see that the SNR has dimensions of inverse of time.
The existence of a characteristic time $\tau$ in our system will
enable us to propose the form of the SNR through the simple scaling
law
\begin{equation}
\label{SNRex}
\mbox{SNR} = f(\alpha,\omega_0\tau)\tau^{-1} \;\; , 
\end{equation}
where $\tau^{-1}=D^{n/(1+n)}\kappa^{1/(1+n)} $ and
$f(\alpha,\omega_0\tau)$ is a dimensionless function, provided that
$v(x)$ does not introduce any other characteristic time, as occurs
when $v(x)=\vert x\vert^\beta$.  We will
suppose that for a given value of $\tau$ the limit of the SNR when
$\omega_0$ goes to zero exists.  As such, the following expression
for small driving frequencies holds
\begin{equation}
\label{SNRap}
\mbox{SNR}= f(\alpha,0)\tau^{-1} . 
\end{equation}

Let us now discuss the main characteristics of this model upon varying
the exponent $n$.  If $n=0$, this system is equivalent to the one
corresponding to Eq. (\ref{model1}), then
one finds the result
$\mbox{SNR}=f(\alpha,\omega_0\kappa^{-1})\kappa$, which does not depend
on the noise
level as shown in Sec. \ref{sec2}.  More interesting is the behavior
obtained for the case
$n>0$.  The scaling of the SNR indicates that it increases when
increasing the noise level, achieving the behavior $\mbox{SNR} \sim
D$ as $n$ goes to infinity.
We note the fact that the only assumption introduced here
concerns to the existence of the SNR in the limit
case of frequency of the external signal going to zero.
A particular and common situation
illustrating this case corresponds to a quartic potential,
obtained when $n=1$, for which the 
SNR increases as $\sqrt D$.

\subsection{Spatially extended systems}
\label{sec3a}

We will now proceed to extend the analysis of Sec. \ref{sec2} to
the case in which spatial effects are considered.
To be explicit, we will treat the
Swift-Hohenberg  (SH)  equation  \cite{cross,swi,ho},
which models  Rayleigh-B\'enard  convection  near the  convective
instability. Far from being specific to this problem, the
main ideas   can also be applied to other
spatially extended systems
in the vicinity of a symmetry breaking bifurcation.

The stochastic  SH equation in dimensionless spatial units is given by
\begin{equation}
\label{SH}
{\partial \psi \over \partial t}=
h(t)\psi -q(1+\nabla^2)^2\psi -g\psi^3
+\xi(\vec r,t) \;\; .
\end{equation}
Here the control parameter $h(t)=-\kappa+\alpha \sin(\omega_0t)$,
with $\kappa$,  $\alpha$, and $\omega_0$  positive constants,
accounts for the
presence of an external  periodic  forcing, due, for instance, to
variations of the  temperature  difference  between the plates of
the convective  cell  \cite{cross,ho,ga2}.  The parameters $g$ and $q$
depend on the  characteristics of the system
and  $\xi(\vec  r,t)$ is Gaussian  white noise with zero mean and
second moment
$\left<\xi(\vec r,t)\xi(\vec r^\prime,t^\prime)\right>
=2D\delta(\vec r - \vec r^\prime)\delta(t-t^\prime)$.

It is interesting to point out the fact that
the spatial effects enter the equation through both  
the term $q(1+\nabla^2)^2\psi$ and the boundary conditions.
Here, however, we will assume that the system is infinite, thus neglecting
the boundary effects. Hence, when dealing with scaling
arguments, the spatial dependence is taken
into account only through the parameter $q$.

With the purpose of analyzing the time evolution  of the system we
will  consider the  convective  heat flux, which in this model is
given by
\begin{equation}
\label{J}
J(t)= c \int \psi(\vec r,t)^2 d \vec r\;\; ,
\end{equation}
where $c$ is a constant depending on the physical characteristics
of the system.  This quantity is in fact the order parameter of
the  transition  from the  homogeneous  state to the state  where
spatial structures  develop.  The existence of spatial ordering can be
revealed by the time-averaged structure factor
\label{S}
\begin{equation}
F(k)=\left<\hat\psi_k\hat\psi_{-k}\right>_t \;\; ,
\end{equation}
where  $\hat\psi_k$ is the spatial Fourier transform of the field
$\psi$ and $\left<\;\right>_t$  indicates  time and noise  average.
The occurrence of a sharp peak in this magnitude  makes the presence
of an ordered spatial structure manifest.

If the noise level is sufficiently small
the field $\psi$ is small, except for a  possible initial  transient, and 
the nonlinear term of Eq. (\ref{SH}) can be neglected. Therefore,
the potential describing the dynamics is scale invariant. 
In this situation, by proceeding as in the
previous section it is
easy to see how the  characteristic  quantities scale with noise.
We note that, when the  linearized  equation is  considered,  any
dimensionless parameter cannot depend on the noise level, because
only $D$  involves  the  dimensions  of the  field  $\psi$.  Thus
$\psi$ scales with the noise as $\psi \sim \sqrt{D}$.  Since the SNR
has dimensions of the inverse of time, then for $J(t)$
it is given by
\begin{equation}
\label{SNRSH}
\mbox{SNR}=\omega_0f_1\left(\gamma \right) \;\; .
\end{equation}
Here $f_1$ is a dimensionless  function which depends on the set
of dimensionless parameters $\kappa/\omega_0$,  $\alpha/\omega_0$
and $q/\omega_0$, denoted by $\gamma$.  We then conclude that the
SNR does not depend on the noise level.  However, both the
output  noise and signal  scale  with the  noise as  $D^2$.  This  fact
indicates that the output signal increases when noise  increases.
In this regard the  structure  factor
also follows a scaling law since $F(k) \sim D$.

\section{Breaking of scale invariance}
\label{sec4}

The class of systems we have discussed in the previous section
are characterized
by the fact that their dynamics exhibit scale invariant potentials.
In this section we will show that scaling arguments can also be
applied when this requirement is not fulfilled.

\subsection{Low noise level}
\label{sec4a}

We will first analyze the case of low noise level.
To be explicit, we will consider the Langevin equation
\begin{equation}
\label{ourmodel3}
{dx \over dt} = -\kappa[1+\alpha\sin(\omega_0 t)]
\left(x+ax^{1+2n}\right) + \xi(t) \;\; . 
\end{equation}
When the noise level is sufficiently small 
the nonlinear term can be neglected.
Then, the SNR does not depend on $D$.
In order to analyze how the SNR behaves upon varying $D$,
we must take into account the nonlinear term.
To this purpose we will assume that the effects of
the nonlinear contribution $ax^{1+2n}$ on a
given quantity, the SNR in this case,
can be replaced by the ones of an 
effective linear term $abD^{n}\kappa^{-n}x$, where
$b \equiv b(\alpha,\omega_0\kappa^{-1})$
is a dimensionless positive function. The explicit form of $b$ may depend
on the quantity we are considering but it is always a positive function.
Consequently the previous equation transforms into
\begin{equation}
\label{ourmodel3b}
{dx \over dt} = -\kappa\left(1+abD^{n}\kappa^{-n}\right)
\left[1+\alpha\sin(\omega_0 t)\right]x
+ \xi(t) \;\; , 
\end{equation}
which can be rewritten in the form
\begin{equation}
\label{ourmodel3c}
{dx \over dt} = -\tilde \kappa[1+\alpha\sin(\omega_0 t)]x
+ \xi(t) \;\; , 
\end{equation}
where $\tilde \kappa=(1+abD^{n}\kappa^{-n})k$ is an effective parameter.
The SNR is then
\begin{equation}
\label{snrmodel3a}
\mbox{SNR} = f(\alpha,\omega_0\tilde \kappa^{-1}) \tilde \kappa
\;\; , 
\end{equation}
which for small frequencies ($\omega_0 \kappa^{-1} \ll 1$) leads to
\begin{equation}
\label{snrmodel3b}
\mbox{SNR} = f(\alpha,0)\kappa\left(1+abD^{n}\kappa^{-n}\right)
\;\; , 
\end{equation}
In this regard, we have been able to predict the behavior of the SNR
as a function of $D$, for low noise level, by means of simple
scaling arguments.
It is interesting to point out the fact that for low noise
level, when $a$ is
positive, the SNR is an increasing function of $D$, whereas
when $a$ is negative the SNR decreases with $D$.
Thus, if the SNR decreases for high noise level, as usually happens, 
the system may exhibit SR when its dynamics around the minimum
of the potential can be approximated by Eq. (\ref{ourmodel3}),
with $a$ positive.

When considering perturbations
to the linear system, the analysis we have  performed
can also be carried out in spatially extended
systems. In order to illustrate this fact, we will now study
the effect of the  nonlinear
term in the SH equation [Eq. (\ref{SH})]. 
Since  for low noise  level the SNR does not  depend  on $D$,
as shown previously, the
lowest order  correction in $D$ to the constant  value of the SNR
comes  from the  nonlinear  term.  To  elucidate  its form, it is
convenient to rewrite Eq.  (\ref{SH}) in the following way
\begin{equation}
\label{SH2}
{\partial \psi \over \partial t}=
\left[-\left(\kappa+g\psi^2\right)+\alpha \sin(\omega_0t)
-q\left(1+\nabla^2\right)^2\right]\psi  +\xi(\vec r,t) \;\; .
\end{equation}
Due to the fact that for low noise level  $\psi^2  \sim D$,
in first  approximation  $\kappa+g\psi^2$  can be
interpreted  as an effective parameter
$\tilde \kappa \equiv  \kappa+g\omega_0^{-1}f_2(\gamma)D$,  with $f_2$
being  a  positive  dimensionless   function.  Consequently,  Eq.
(\ref{SH2}) has the same form as the linearized counterpart for which
the scaling law for the SNR [Eq.  (\ref{SNRSH})]  has been derived.
We note that this case differs from the previous one in the form
in which the nonlinearity enters the equation, therefore we must
follow an alternative scheme.
By replacing  $\kappa$ by $\tilde \kappa$ in Eq.  (\ref{SNRSH}) and
expanding around $\kappa$ we then obtain
\begin{equation}
\label{SNRSH2}
\mbox{SNR} = \omega_0f_1\left(\gamma \right)
+ { \partial f_1 \over \partial \kappa}
g\omega_0^{-1}f_2(\gamma)D  \;\; ,
\end{equation}
which includes the lowest order  correction in $D$ to the SNR due
to the nonlinear  term.  An important  consequence of this result
is the fact that the  knowledge of the  dependence  of the SNR on
the parameter  $\kappa$, for the linearized  equation, enables us
to  predict  the  presence  of SR  when  the  nonlinear  term  is
considered.  If $f_1$ is an increasing function of $\kappa$ then
$\partial  f_1 / \partial  \kappa $ is positive and  consequently
the SNR is an  increasing  function  of $D$ for low noise  level.
Since for high $D$ the SNR decreases, one then concludes  that it
exhibits a maximum thus indicating the presence of SR.
In fact, it has been shown that the SH equation may
exhibit SR since for a certain range
of the values of the parameter $\partial  f_1 / \partial  \kappa $
is positive \cite{prl3}. We would like to emphasize
the fact that the previous
scaling  argument is robust upon varying the nonlinear term.  For
example, if one replaces the term $g\psi^3$ by $g|\psi|\psi$  one
obtains a SNR that  increases  as  $\sqrt{D}$  instead of as $D$.

\subsection{High noise level}
\label{sec4b}

As in the former case, the high noise level limit can also
be treated by means of scaling arguments. 
To this purpose we will assume that the dynamics of the
system in this case may be approximated  by
\begin{equation}
\label{ourmod2}
{dx \over dt} = -\kappa[1+\alpha\sin(\omega_0 t)]x^{n} -lx^{m}
+ \xi(t) \;\; , 
\end{equation}
where $l$, $n$, and $m$ are positive constants.
If $n=m$, Eq. (\ref{ourmod2}) is equivalent to Eq. (\ref{ourmodel}),
as follows by
only changing the values of the parameters. If
$n>m$, Eq. (\ref{ourmod2}) also leads to Eq. (\ref{ourmodel}),
since when $n>m$, for high noise level, the term $lx^{m}$
can be neglected.
Therefore we consider the case in which $n<m$.

The previous equation can be rewritten
in the following way:
\begin{equation}
\label{ourmod2b}
{dx \over dt} = -l\left[1+ {\kappa\over lx^{m-n}}+
{\kappa\alpha \over l x^{m-n}}
\sin(\omega_0 t)\right]x^{m} + \xi(t) \;\; .
\end{equation}
Since for high noise level $x$ is large ($x \sim D^{1/(1+m)}$)
the periodic force acts as a small perturbation
to the dynamics of the system . Thereby,
proceeding in a similar way as in the case for low
noise level, one can introduce the effective parameter
$\tilde \alpha \equiv bl^{-(n+1)/(1+m)}\kappa
D^{-(m-n)/(1+m)}\alpha$, with $b$  now a dimensionless constant.
Since the term $\kappa/(lx^{m-n})$ can be neglected,
Eq. (\ref{ourmod2b}) reads
\begin{equation}
\label{ourmod2c}
{dx \over dt} = -l\left[1+ \tilde \alpha\sin(\omega_0 t)\right]x^{m}
+ \xi(t) \;\; , 
\end{equation}
which has the same form as  Eq. (\ref{ourmodel}). We then obtain
\begin{equation}
\mbox{SNR} = f(\tilde\alpha,\omega_0\tau)\tau^{-1} \;\; , 
\end{equation}
with $\tau=D^{-(m-1)/(1+m)}l^{-2/(1+m)}$.
Since $\tilde \alpha$  and $\omega_0\tau$ are small for
high $D$, then
\begin{equation}
\mbox{SNR} = {1 \over 2}f^{\prime\prime}{\tilde \alpha}^2\tau^{-1} \;\; , 
\end{equation}
where $f^{\prime\prime}$ is the second derivative of
$f(\tilde\alpha,\omega_0\tau)$, with respect to $\tilde\alpha$, evaluated
at $\tilde\alpha=0$ and $\omega_0\tau=0$.
Explicitly,
\begin{equation}
\mbox{SNR} = {1 \over 2}f^{\prime\prime}\alpha^2\kappa^2b^2
l^{-2n / (1+m)}D^{-1 + 2n/ (1+m)}\;\; . 
\end{equation}
Note that when the forcing term does not depend on $x$, i. e. $n=0$,
the SNR
always decreases as $\mbox{SNR} \sim D^{-1}$, irrespective of the
value of $m$.
From this expression one can elucidate some interesting situations.
For instance if $m=2n-1$, the SNR tends to a constant value for
high noise level, whereas if $m<2n-1$ it diverges. 
Hence, in this situation, for $m<2n-1$ the response of the system
is always enhanced when the noise level is increased. Thus,
noise is unable to destroy the coherent response of the system
to the periodic input signal.

\section{Discrete symmetries}
\label{sec5}

In the cases discussed previously, we have considered the invariance of
the system
under a continuous scaling of the noise level.
Another interesting possibility is that the system
may only be invariant for a discrete set of values of the
noise level. This fact implies that the quantities which
remain invariant under that transformation exhibit
a periodic behavior with the logarithm of the noise level.

In order to treat this aspect explicitly
we now consider 
the following Langevin dynamics
\begin{equation} \label{model}
{dx \over dt} = -F(x,t)x +\sqrt{2D}\xi(t) \;\; ,
\end{equation}
where $F(x,t)$ is a given function, $\xi(t)$ is Gaussian white noise
with zero mean and second moment
$\left<\xi(t)\xi(t+\tau)\right>=\delta(\tau)$, and $D$ is a
constant, defining the noise level.  Here the input signal enters the
system through $F(x,t)$, and we will assume it to be periodic in
time with frequency $\omega_0/2\pi$.  The output of the system is given
by $v(x)=|x|^n$, with $n$ a positive constant.

The transformations
\begin{eqnarray}
x &\rightarrow x^{\prime} \equiv &e^\gamma x \;, \nonumber \\
D &\rightarrow D^{\prime} \equiv &e^{2\gamma} D \;,
\end{eqnarray}
with $\gamma$ a constant leaves
Eq.  (\ref{model}) and the SNR [Eq. (\ref{SNRdef})]
invariant, provided that
\begin{equation} \label{master}
F(x,t)=F(xe^\gamma,t) \;\; .
\end{equation}
As a consequence, for the class of systems for which the previous
requirement holds, for a certain value of $\gamma$, the SNR has the
same value at D and at $e^{2\gamma}D$.  This fact occurs when
$F(x,t)=q[\ln(x),t]$, where $q$ is a periodic function of its first
argument, with periodicity $\gamma$, if $\gamma$ is the lower
positive number satisfying Eq.  (\ref{master}).  Therefore, the SNR
is a periodic function of the logarithm of the noise level.
We note 
that both  signal and noise are not invariant under this
transformation, contrarily they change as
\begin{eqnarray}
S^{\prime}&=&e^{2\gamma n}S
\;\;, \nonumber \\
Q^{\prime}&=&e^{2\gamma n}Q
\label{SandN}
\;\;.
\end{eqnarray}

\begin{figure}[th]

\centerline{
\epsfxsize=7cm 
\epsffile{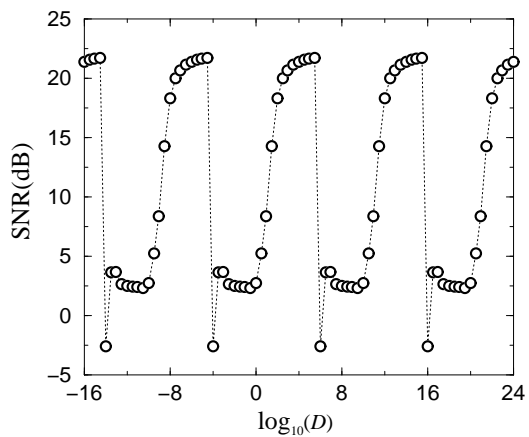}}

\caption[b]{\label{fig1}
SNR corresponding to $F(x,t)$ given through Eq.  (\ref{ej1})
with parameter values $\kappa_1=100$, $\kappa_2=1$, $T=10$,
$\alpha=0.5$ and $\omega_0/2\pi=1$.
}
\end{figure}

In order to illustrate the previous results we have analyzed a
representative explicit expression of $F(x,t)$.
The corresponding Langevin equation has been numerically integrated  by
means of a standard second-order Runge-Kutta method for stochastic
differential equations \cite{kloeden}.
As the output of
the system we have used $v(x)=x^2$.  We will consider the case
in which
\begin{equation} \label{ej1}
F(x,t)=\Theta_T[\log_{10}(x^2)][1+\alpha\cos(\omega_0 t)] \;\; ,
\end{equation}
where $\alpha$ and $\omega_0$ are constants, and $\Theta_T(s)$ is a
square wave of period $T$ defined by
\begin{equation}
\Theta_T(s)=\left\{
\begin{array}{lcl}
\kappa_1 &  & \mbox{if} \; \sin(2\pi  s /T) > 0 \;\; , \nonumber \\
\kappa_2 &  & \mbox{if} \; \sin(2\pi s /T) \le 0 \;\; ,
\end{array}
\right.
\end{equation}
with $\kappa_1$ and $\kappa_2$ constants.  In Fig.  \ref{fig1} we have
plotted the SNR corresponding to the previous form of $F(x,t)$, for
particular values of the parameters.  This figure clearly manifests
the periodicity of the SNR as a function of the noise level and the
presence of multiple maxima at $D=D_0e^{mT}$, with $m$ being any
integer number and $D_0$ the noise level corresponding to the
maximum with $m=0$.
The appearance of multiple maxima then implies
the presence of  stochastic multiresonance \cite{prl2}.

From this example one can infer the mechanism responsible for the
appearance of this phenomenon.  Due to the fact that the SNR has
dimensions of the inverse of time, its behavior is
closely related to the characteristic temporal scales of the system.
Thus variations of the relaxation time manifest in the SNR.  In this
example, when $T$ is sufficiently large, for some values of the
noise level the system may be approximated by
\begin{equation}\label{ourmodel5}
{dx \over dt} = -\kappa_i[1+\alpha\sin(\omega_0 t)]x +
\sqrt{2D}\xi(t) \;\; ,
\end{equation}
where $i=1,2$, depending on the noise level.
In such a situation the SNR is given by
\begin{equation}\label{SNRex2}
\mbox{SNR} = f(\alpha,\omega_0\kappa_i^{-1})\kappa_i \;\; ,
\end{equation}
with $f$ a dimensionless function.  For a sufficiently
low frequency, the SNR is proportional to $\kappa_i$ [$\mbox{SNR} =
f(\alpha,0)\kappa_i$], i.e.  proportional to the inverse of the
relaxation time.  Consequently, there are two set of values of $D$
for which the SNR differs in approximately $10\log_{10}(\kappa_1/\kappa_2)$
dB, as one can see in Fig.  \ref{fig1}.  We then conclude that
multiple maxima in the SNR appears as a consequence of the form in
which the relaxation time of the system changes with the noise
level.

\section{Concluding remarks}
\label{sec6}

In this paper we have shown how scaling arguments can be used to
derive the main characteristics of a broad variety
of periodically modulated noisy systems, focusing
on the phenomenon of stochastic resonance.
Application of scaling arguments then leads to the knowledge
of the dependence of the signal-to-noise ratio on
the noise level. In particular, we have shown
that the signal-to-noise ratio may increase when the noise
level is increased.
This fact makes the presence of stochastic resonance manifest, implying
that the addition of noise enhance the response of
the system to a periodic signal.
In this context, 
the constructive role played by
noise, sometimes considered counterintuitive, under some
circumstances is merely a consequence of the form
in which the system scales upon variations of 
the noise level and may arise directly from dimensional
analysis.

Finally, it is worth pointing out the fact that
in this paper we have presented explicit results concerning to
representative situations of interest. However, the methodology we have
outlined can be applied to  a broad variety of
different situations
due to the general assumptions involved in the scaling arguments.
Our results, then, reinforce
the usefulness of scaling concepts in this field, adding
new examples to the already wide variety of cases encountered
in Statistical Physics.

\section*{Acknowledgments}

This work was
supported by DGICYT of the Spanish Government under Grant No.
PB95-0881.
J. M. G. V. wishes to thank Generalitat
de Catalunya for financial support.


\begin{references}
\bibitem{dipeq}
The  relaxational  dynamics of a single  dipole in an
external  field   may be  described  by the  Langevin  equation
governing  the  dynamics  of the  unit  vector  $\hat  R$  along  the
direction of the dipole moment
%
$$
{d \hat R \over dt} =
\left[\vec h(t)
\times \hat R + \vec \xi(t) \right] \times \hat R \;\; .
$$
%
Here the total field  $\vec  h(t)=k(1+\alpha\sin(\omega_0t))\hat  z$,
with  $\hat  z$  a  unit  vector,  consists  of a  constant  plus  an
oscillating  field and is characterized  by the parameters $k$ and
$\alpha$.  The  noise  term  $\vec  \xi(t)$  is Gaussian and  white,
with zero mean and second moment
$\left<\xi_i(t)\xi_j(t+\tau)\right>=D\delta_{ij}\delta(\tau)$.
\end{references}
\end{document}